\documentclass{IEEEtran4PSCC}

\ifCLASSINFOpdf
   \usepackage[pdftex]{graphicx}
\else
   \usepackage[dvips]{graphicx}
\fi

\usepackage{amsthm}
\usepackage{amsmath}
\usepackage{amsfonts}
\usepackage{algpseudocode}
\usepackage{algorithm}
\usepackage[table]{xcolor}
\usepackage{multirow}
\usepackage{pgfplots}
\newcommand{\bs}[1]{\boldsymbol{#1}}

\makeatletter
\let\old@ps@headings\ps@headings
\let\old@ps@IEEEtitlepagestyle\ps@IEEEtitlepagestyle
\def\psccfooter#1{%
    \def\ps@headings{%
        \old@ps@headings%
        \def\@oddfoot{\strut\hfill#1\hfill\strut}%
        \def\@evenfoot{\strut\hfill#1\hfill\strut}%
    }%
    \def\ps@IEEEtitlepagestyle{%
        \old@ps@IEEEtitlepagestyle%
        \def\@oddfoot{\strut\hfill#1\hfill\strut}%
        \def\@evenfoot{\strut\hfill#1\hfill\strut}%
    }%
    \ps@headings%
}
\makeatother

\psccfooter{%
        \parbox{\textwidth}{\hrulefill \\ \small{23nd Power Systems Computation Conference} \hfill \begin{minipage}{0.2\textwidth}\centering \vspace*{4pt} \includegraphics[scale=0.06]{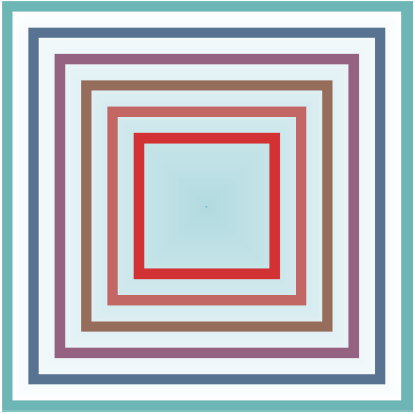}\\\small{PSCC 2024} \end{minipage} \hfill \small{Paris-Saclay, France --- June 4 -- June 7, 2024}}%
}

\pgfplotsset{compat=1.18}
\begin{document}

\title{Certification of MPC-based zonal controller security properties using accuracy-aware machine learning proxies}

\author{
\IEEEauthorblockN{Pierre HOUDOUIN, Manuel RUIZ, Lucas SALUDJIAN, Patrick PANCIATICI}
\IEEEauthorblockA{French Transmission System Operator, RTE, Paris, France\\
\{pierre.houdouin, manuel.ruiz, patrick.panciatici\}@rte-france.com}}

\maketitle

\thanksto{\noindent Submitted to the 23nd Power Systems Computation Conference (PSCC 2024).}

\begin{abstract}
    The fast growth of renewable energies increases the power congestion risk. To address this issue, the French Transmission System Operator (RTE) has developed closed-loop controllers to handle congestion. RTE wishes to estimate the probability that the controllers ensure the equipment's safety to guarantee their proper functioning. The naive approach to estimating this probability relies on simulating many randomly drawn scenarios and then using all the outcomes to build a confidence interval around the probability. Although theory ensures convergence, the computational cost of power system simulations makes such a process intractable.

    The present paper aims to propose a faster process using machine-learning-based proxies. The amount of required simulations is significantly reduced thanks to an accuracy-aware proxy built with Multivariate Gaussian Processes. However, using a proxy instead of the simulator adds uncertainty to the outcomes. An adaptation of the Central Limit Theorem is thus proposed to include the uncertainty of the outcomes predicted with the proxy into the confidence interval. As a case study, we designed a simple simulator that was tested on a small network. Results show that the proxy learns to approximate the simulator's answer accurately, allowing a significant time gain for the machine-learning-based process.
\end{abstract}


\section{Introduction}

Integrating renewable energies on a large scale poses challenges in the operation and management of power systems. It leads to unpredictable and variable flow injections into transmission lines, increasing the risk of power congestion. To address these challenges, the French Transmission System Operator (TSO), RTE, has adopted a decentralized management approach, dividing the entire system into sub-transmission areas (zones). Real-time constraints within each zone are managed through a local closed-loop controller called NAZA \cite{Hoang2021Power}, \cite{Hoang2021Predictive}, \cite{Straub2018Zonal}. Designed to handle local problems with local actions, it enables the management of battery devices, topological modifications, and curtailment of renewable production \cite{Meyer2020Power} inside the zone. Alongside the massive deployment of these controllers across the network, RTE aims to obtain guarantees regarding their proper functioning. Given a set of renewable power production scenarios, RTE seeks to compute the probability of NAZA ensuring the equipment's safety, referred to as $\mathbf{p_{safe}}$ thereafter. Equipment's safety is ensured if congestions are avoided during the scenario simulation. If any line in the zone becomes overloaded during the simulation, it is considered a security threat.

\vspace{0.2cm}

The estimation of $\mathbf{p_{safe}}$ relies on the law of large numbers. The most basic approach is the brute-force process: scenarios are randomly drawn and simulated until $\mathbf{p_{safe}}$ is accurately estimated. Central Limit Theorem (CLT) \cite{Fisher2011History} ensures the convergence of this process. The more iterations are performed, the more accurate the estimation becomes \cite{Jeffrey1998Exact}. In our case, only extreme and thus rare-to-observe scenarios will likely pose a security threat. A large number of scenarios must, therefore, be drawn to observe enough threatening situations and obtain a reliable estimation of the probability. Although it theoretically works, the drawback of the brute-force process is that it requires a lot of simulations. As power system simulations are computationally costly, such a process is intractable in this case. However, the outcome of many scenarios can be figured out without a simulation. For instance, a scenario without wind will surely not lead to congestion if the network is correctly exploited. Hence, learning a proxy of the simulator to avoid unnecessary simulations and focus only on the interesting ones can lead to considerable time gain.

\vspace{0.2cm}

The use of proxies to reduce global computational costs meets a growing interest in all domains of power systems. In 2016, Canyasse et al. \cite{Canyasse2017Supervised} used supervised learning algorithms to build real-time proxies for solving Alternative Current Optimal Power Flow. Two years later, Duchesne \cite{Duchesne2018Using} investigated machine-learning proxies to deal with Security Constraint Optimal Power Flow in the context of operation planning. Also, in 2018, Dalal et al. \cite{Dalal2018Unit}, \cite{Dalal2019Chance} considered the Nearest Neighbors algorithm to predict short-term decision outcomes and applied it to the outage scheduling problem. More recently, in 2022, Chen et al. \cite{Chen2022Learning} proposed a deep-learning-based proxy to solve the Security Constraint Economic Dispatch problem and handle real-time applications efficiently. 

\vspace{0.2cm}

In this paper, we will also use a proxy to address the computational limitations of the brute-force approach. Our new proxy-based process aims to avoid useless simulations and achieve a faster estimation of $\mathbf{p_{safe}}$. Using batches of simulations, we train a \textbf{multivariate Gaussian process (MGP)} \cite{Rasmussen2003Gaussian} to predict the scenario outcome. \textbf{Gaussian process (GP)} also became very popular in the power systems community \cite{Kocijan2016Modelling}, \cite{Jain2018Learning}, \cite{Yang2018Power}. Indeed, they provide exact confidence intervals around the prediction and enable the incorporation of prior knowledge of the system into the proxy \cite{Ospina2011Learning}. Parameters of the conditional distribution are continuously updated with the new batch of data to enhance its accuracy throughout the process. The confidence interval provided with the GP's prediction enables permanent assessment of the prediction's quality. If the uncertainty in the prediction of the scenario outcome is acceptable, we keep the prediction and avoid a simulation. Otherwise, we perform a simulation. This leads to significantly sped-up iterations when the simulation is not performed. Finally, the CLT is adapted using the Lyapunov version to include the uncertainty of the proxy's predictions in the confidence interval of $\mathbf{p_{safe}}$'s estimation. 

\vspace{0.2cm}

The objective of this work is two-fold:
\begin{itemize}
    \item Train an accurate proxy of a black-box simulator using an MGP
    \item Show that with an accurate proxy, the machine-learning-based process converges much faster toward $\mathbf{p_{safe}}$ for a given precision, with even a greater gain for demanding precision
\end{itemize}

The article's structure is as follows: Section II introduces the notations, reminders on the MGP, and problem formulation. Section III presents both the brute-force and the proxy-base processes. Section IV provides the computational results of the case study. Conclusions and future perspectives are discussed in Section V.

\section{Preliminaries and problem formulation}

\subsection{Notations}

Throughout this paper, upper-case (lower-case) boldface letters will be used for matrices (column vectors), and $(.)^T$ denotes the transposition.
Given a zone of the electric network, we use the following notations :

\begin{itemize}
	\item $\mathcal{Z}_L=\{L_1,...,L_L\}$ is the set of lines in the considered zone, $L$ is its cardinality
	\item $\mathcal{Z}_N=\{N_1,...,N_N\}$ is the set of nodes in the considered zone, $N$ is its cardinality
	\item $P_n$ is the allowed renewable power injection at node $N_n$
 	\item $PA_n$ is the available renewable power injection at node $N_n$
    \item $P_n^{max}$ is the maximum renewable power that can be produced at node $N_n$, and $\frac{PA_n}{P_n^{max}} \in [0,1]$ is the relative available renewable power injection
	\item $F_l$ and $\bar{F_l}$ are respectively the power flow and the IST of line $L_l$
\end{itemize}

\subsection{Multivariate gaussian process}

Consider an unknown function $h: \mathbb{R}^N \rightarrow \mathbb{R}^L$ that represents the non-deterministic answer $\bs{y} \in \mathbb{R}^L$ of a system to an input $\bs{x} \in \mathbb{R}^N$. We model $\bs{y} = h(\bs{x}) + \mathcal{N}(0, \sigma_0^2 \bs{I})$. Multivariate Gaussian process \cite{Chen2019Multivariate} (MGP) regression aims to learn the underlying dynamic of $h$ by supposing that it is the realization of a multivariate stochastic Gaussian process. The following elements fully characterize the multivariate stochastic process:

\begin{itemize}
	\item Its mean function : $\mu : \mathbb{R}^N \rightarrow \mathbb{R}^L$
	\item Its correlation (kernel) function : $k : \mathbb{R}^N \times \mathbb{R}^N \rightarrow \mathbb{R}$
	\item Its output covariance matrix : $\bs{\Omega}$, a $L \times L$ matrix
\end{itemize}

Prior knowledge of the system's dynamic can be incorporated into these three elements. For a given $\bs{x} \in \mathbb{R}^N$, $\mu(\bs{x})$ represents the prior expected value of $h(\bs{x})$. Usually, as no specific information about $h$ is known, the choice is $\mu = 0$, which is also our choice here. The kernel function $k(\bs{x_1},\bs{x_2})$ measures to what extent $\bs{y_1}$ and $\bs{y_2}$, the outputs of $\bs{x_1}$ and $\bs{x_2}$, are correlated. Its selection has to be adapted to the problem. For our problem, we choose the stationary squared exponential kernel function : 

\begin{align*}
    k(\bs{x_1}, \bs{x_2}) = \sigma_f^2 \exp \left( - \frac{|\bs{x_1}-\bs{x_2}|^2}{2l^2} \right) + \sigma_0^2 \delta(\bs{x_1}, \bs{x_2})
\end{align*}

It is a parametric function with parameters $\bs{\theta} = \left[ \sigma_0^2, \sigma_f^2, l \right]$. Parameter $\sigma_0^2$ represents the variance of the system's answer to the same input, $\sigma_f^2$ reflects the signal variance, and $l$ corresponds to the characteristic length-scale of the kernel. The optimal parameters are learned by maximizing the likelihood \cite{Rasmussen2003Gaussian}, \cite{Duvenaud2014Automatic}, \cite{Petit2023Parameter}. The squared exponential kernel results in a smooth prior on $h$ and reflects that similar scenarios are expected to have similar threat potential for the network. Knowing one scenario's outcome only provides local information about neighbors' scenarios' outcomes. The correlation between their output indeed decreases exponentially with the distance. Finally, the process's output covariance matrix reflects the correlations between the output's coordinates. This matrix allows the posterior distribution of the MGP to preserve the covariance structure of the output. The choice of $\bs{\Omega} = I_{L}$ implies that coordinates are decorrelated. The MGP then boils down to training $L$ separate univariate GP to predict each coordinate independently. In this work, $\bs{\Omega}$ is fixed at the beginning of the process. It represents the line's maximum flow correlation and is chosen appropriately regarding the network's structure. An extension of this work to integrate an estimation of $\bs{\Omega}$ across the process's iterations is planned. We now define the following variables :

\begin{itemize}
	\item $\bs{X_m} = \left[\bs{x_1},...,\bs{x_m}\right]$ the inputs vectors
	\item $\bs{Y_m} = \left[\bs{y_1}^T,...,\bs{y_m}^T\right]$ the corresponding simulated outputs
	\item $\bs{M_m} = \left[\mu(\bs{x_1}),...,\mu(\bs{x_m})\right]$ the corresponding mean vectors
	\item $\bs{\Sigma_m} = \left( k(\bs{x_i}, \bs{x_j}) \right)_{i,j \in [1,m]}$ the covariance matrix between the input vectors
	\item $\bs{x_{m+1}}$ a new input vector
	\item $\bs{\Sigma_{[1, m], m+1}} = \left[ k(\bs{x_1}, \bs{x_{m+1}}),...,k(\bs{x_m}, \bs{x_{m+1}}) \right]$ the covariance vector between the new input and the old ones
\end{itemize}

The joint distribution of $\bs{Y_{m+1}}$ is a matrix-variate gaussian distribution \cite{Mathai2022Multivariate} $\mathcal{M} \mathcal{N} \left( \bs{M_{m+1}}, \bs{\Sigma_{m+1}}, \bs{\Omega} \right)$ \cite{Chen2019Multivariate}. Similarly to the univariate GP, we can derive the conditional distribution of $\bs{y_{m+1}} | \bs{y_1},...,\bs{y_m}$ : 

\begin{align*}
    &\bs{\mu_*} = \mu(\bs{x_{m+1}}) + \bs{\Sigma_{[1, m], m+1}}^T \bs{\Sigma_m}^{-1}(\bs{Y_m}-\bs{M_m}) \\
    &\sigma_* = k \left(\bs{x_{m+1}}, \bs{x_{m+1}}\right) - \bs{\Sigma_{[1, m], m+1}}^T \bs{\Sigma_m}^{-1} \bs{\Sigma_{[1, m], m+1}}^T \\
    &\bs{y_{m+1}} \in \mathbb{R}^L \sim \mathcal{N} \left( \bs{\mu_*}, \sigma_* \bs{\Omega} \right)
\end{align*}

The obtained multivariate conditional distribution is very similar to the univariate case. The observed points matter for the mean $\bs{\mu_*}$ and the covariance matrix's scale  $\sigma_*$. The covariance matrix's shape is always $\bs{\Omega}$, which ensures that the conditional distribution preserves the correlations between the output coordinates.

\subsection{Problem formulation}

Given : 

\begin{itemize}
    \item A zone of the power network
    \item A black-box simulator $\mathcal{S}$ to simulate the zone's behavior in answer to power injections
    \item A set of renewable power injection scenarios $\mathcal{X}$
\end{itemize}

, we aim to estimate the probability $\mathbf{p_{safe}}$ to draw a scenario $\bs{x} \in \mathcal{X}$ well-handled by NAZA, i.e., that do not lead to overloaded lines.

\vspace{0.2cm}

\underline{\textbf{Hypothesis on the zone}}

The zone is composed of $L$ lines, $N$ generator nodes, and a NAZA controller to handle congestion management. We consider the following working hypothesis : 
\begin{enumerate}
    \item NAZA controller can curtail each node's renewable energy power injections, thus reducing $P_n \leq PA_n$.
    \item Each generator produces the maximum available or allowed power
    \item The loads are constants
    \item DC approximation: we neglect the reactive power
\end{enumerate}

\vspace{0.2cm}

\underline{\textbf{Renewable energy generation scenario}}

We model a renewable energy production scenario $\bs{x}$ by the vector of the $N$ available renewable power injections at each network node. To facilitate MGP learning, we draw the relative available renewable power injection. Our set of scenarios $\mathcal{X}$ then boils down to $[0, 1]^N$, which is easier to explore and will be referred to as the \textit{information structure}. To preserve the spatial correlation of the nodes, the power injection vector $\bs{x}$ is generated using a truncated multivariate Gaussian distribution.

\begin{align*}
\bs{x} = \begin{pmatrix} x_1 \\ . \\ . \\ . \\ x_N \end{pmatrix} \sim \mathcal{N}_{|\mathcal{X}}\left(0, \bs{\Sigma_{nodes}} \right) \longrightarrow \begin{pmatrix} PA_1 = x_1 \times P^{max}_1 \\ . \\ . \\ . \\ PA_N = x_N \times P^{max}_N \end{pmatrix}
\end{align*}

We consider here a very restrictive subspace of low-dimensional scenarios. Our main goal is to show how an MGP can learn the multivariate answer of a black-box function. Future work will deal with more complex scenarios, including a temporal aspect, to simulate the network's behavior on many time steps.

\underline{\textbf{Network simulation}}

The network's evolution is simulated using a \textit{network simulator} $\mathcal{S}$. It takes as input a renewable power production scenario $\bs{x}$ and returns the vector of the lines' relative power flow $\bs{y} \in \mathbb{R}^{L}$. For line $l$, the relative power flow is $\bs{y}^{(l)} = \frac{F_l}{\bar{F_l}}$. The \textit{network simulator} used here is a simple simulator implemented by the authors (more details are provided in the experiments section). Again, an extension of this work, on a real zone of the French transmission network and using a real network simulator \cite{Marin2022Open}, is planned.

\vspace{0.2cm}
 
\underline{\textbf{Well-handled scenarios}}

A scenario $\bs{x}$ is considered well-handled by NAZA if no congestions on any lines occur. Line $L_l$ is considered congested if its power flow exceeds the IST. Thus, denoting by $f$ the Boolean function that indicates whether a scenario is well-handled or represents a security threat, we have :

\begin{align*}
f\left(\bs{y}\right) = z = \left\{
    \begin{array}{ll}
        1 & \mbox{if } \forall l \in [1, L], \bs{y}^{(l)} <= 1 \\
        0 & \mbox{if } \exists l \in [1, L], \bs{y}^{(l)} > 1
    \end{array}
\right.
\end{align*}

\begin{figure}[H]
\centering
\begin{tikzpicture}
\node[draw] at (0,0) {$\begin{pmatrix} x_1 \\ . \\ . \\ . \\ x_N \end{pmatrix}$};
\node[draw] at (2.7,0) {$\begin{pmatrix} P_1^{max} \\ . \\ . \\ . \\ P_N^{max} \end{pmatrix}$};
\node[draw] at (6,0) {$\begin{pmatrix} PA_1 = x_1 \times P^{max}_1 \\ . \\ . \\ . \\ PA_N = x_N \times P^{max}_N \end{pmatrix}$};
\node[draw,align=center] at (6,-3.2) {\quad Simulator $\mathcal{S}$ \quad \quad \quad \\ \\ Zone $\mathcal{Z}$ \\ \\ Lines features};
\draw[-] (6.8,-3.32)--(7.3,-3.32);
\draw[-] (6.8,-3.32)--(7.1,-3.1);
\draw[-] (7.3,-3.32)--(7.1,-3.1);
\fill (6.8,-3.32) circle[radius=2pt];
\fill (7.3,-3.32) circle[radius=2pt];
\fill (7.1,-3.1)    circle[radius=2pt];
\node[draw] at (2.6,-3.2) {$\begin{pmatrix} y_1 \\ . \\ . \\ . \\ y_L \end{pmatrix}$};
\draw[red, font=\large] (4.5,-1.6) node {\textbf{Black-box}};
\draw[red, thick,dashed] (1.5,-4.7) -- (8.1,-4.7) -- (8.1,1.5) -- (1.5,1.5) -- (1.5,-4.7);
\draw[->,ultra thick] (0.65,0)--(1.85,0) node{};
\draw[->,ultra thick] (3.55,0)--(4.07,0) node{};
\draw[->,ultra thick] (6,-1.18)--(6,-2.13) node{};
\draw[->,ultra thick] (4.41,-3.2)--(3.22,-3.2) node{};
\node[draw,align=center] at (0.2,-3.2) {$z \in \{0, 1\}$};
\draw[->,ultra thick] (1.98,-3.2)--(1.05,-3.2) node{};
\draw (1.7,-3.6) node {$f$};
\end{tikzpicture}
\caption{Black-box function taking as input a renewable energy generation scenario $\bs{x}$ and returning $0$ if a congestion occurred, $1$ if security was ensured}
\label{fig:1} 
\end{figure}
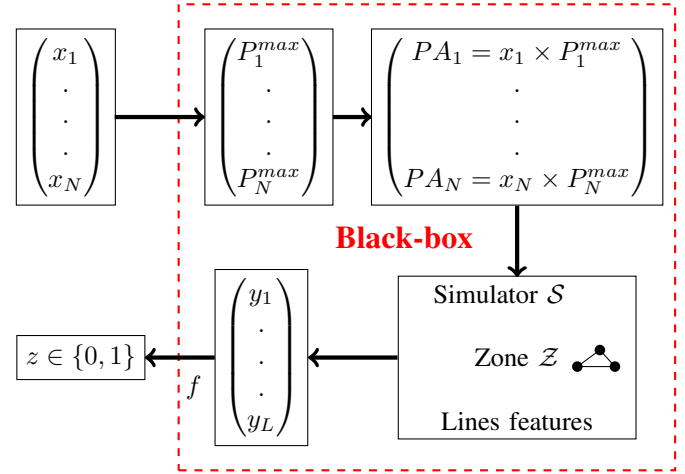 

The zone, the nodes, lines features, and the simulator constitute a \textit{black-box function}, as illustrated in Fig.1. The goal of this work is to estimate as precisely and as fast as possible the probability for the black-box function to return $1$: $\mathbf{p_{safe}} = \mathbb{P} \left( f \circ \mathcal{S} (\bs{x}) = 1 \right)$. The following section will detail two processes to estimate $\mathbf{p_{safe}}$. 

\section{Processes presentation}

\subsection{Brute-force process}

The brute-force process is the vanilla way to estimate $\mathbf{p_{safe}}$. We randomly sample scenarios in $\mathcal{X}$, compute $f \circ \mathcal{S}$ using the network simulator, and use the CLT to build a confidence interval around $\mathbf{p_{safe}}$. We iterate the sampling until a satisfying confidence interval is reached. Given the $m$ simulation results $z_1,...,z_m$, $\overline{z_m} = \frac{1}{m} \sum_{i=1}^m z_i$ their average, $Q_{\alpha}$ the $1-\alpha$ quantile of the standard normal distribution, the asymptotic confidence interval [6] of level $1-\alpha$ around $\mathbf{p_{safe}}$ is :

\begin{align*}
I_{\alpha} &= [p_{min}, p_{max}]\\
p_{min} &= \frac{\overline{z_m} + \frac{Q_{\frac{\alpha}{2}}^2}{2m} - Q{\frac{\alpha}{2}} \sqrt{ \frac{\overline{z_m}\left(1-\overline{z_m}\right)}{m} + \frac{q_{\frac{\alpha}{2}}^2}{4m^2}}}{1 + \frac{Q_{\frac{\alpha}{2}}^2}{m}} \\
p_{max} &= \frac{\overline{z_m} + \frac{Q_{\frac{\alpha}{2}}^2}{2m} + Q_{\frac{\alpha}{2}} \sqrt{ \frac{\overline{z_m}\left(1-\overline{z_m}\right)}{m} + \frac{q_{\frac{\alpha}{2}}^2}{4m^2}}}{1 + \frac{Q_{\frac{\alpha}{2}}^2}{m}} \\
\end{align*}

The brute-force process is summarized in Algorithm 1.

\begin{algorithm}[!t]
\caption{Brute-force process} \label{alg:CV1}
\textbf{Inputs} : 
\begin{itemize}
    \item scenario sampler $\mathbb{X}$
    \item network simulator $\mathcal{S}$
    \item confidence interval precise enough : $CI_{good}?$ tool
    \item confidence interval level $1 - \alpha$
\end{itemize}
\textbf{Output} : $I = [p_{min}, p_{max}]$, $1 - \alpha$ confidence interval level around $\mathbf{p_{safe}}$
\begin{algorithmic}[1]
\State Initialize $m=0$ and $I_m = [0, 1]$
\State Initialize the bag of outputs $\mathcal{B} = \{\}$
\While{not $CI_{good}?(I_m)$}
    \State Draw a scenario $\bs{x_m}$ using $\mathbb{X}$
    \State Increase $m=m+1$
    \State Compute the outcome $z_m = f \circ \mathcal{S}(\bs{x_m})$
    \State Add outcome $z_m$ to the bag $\mathcal{B} = \mathcal{B} \cup \{z_m\}$
    \State Compute $I_m$ of level $1 - \alpha$ using $\mathcal{B}$
\EndWhile
\end{algorithmic}
\end{algorithm}

\subsection{Proxy-based process}

The issue with the brute-force process is the very high amount of required simulations to get an accurate estimation of $\mathbf{p_{safe}}$, especially if $\mathbf{p_{safe}}$ is close to 0 or 1. Fig.2 depicts the number of simulations required for the brute-force process to reach the desired relative precision on the estimated probability for different values of $\mathbf{p_{safe}}$. To reduce the number of required simulations, we propose to learn the response $\bs{y} = \mathcal{S}(\bs{x})$ of the simulator to a scenario. We train an MGP-based proxy $\Tilde{\mathcal{S}}$ with the previously simulated scenarios.

\vspace{0.2cm}

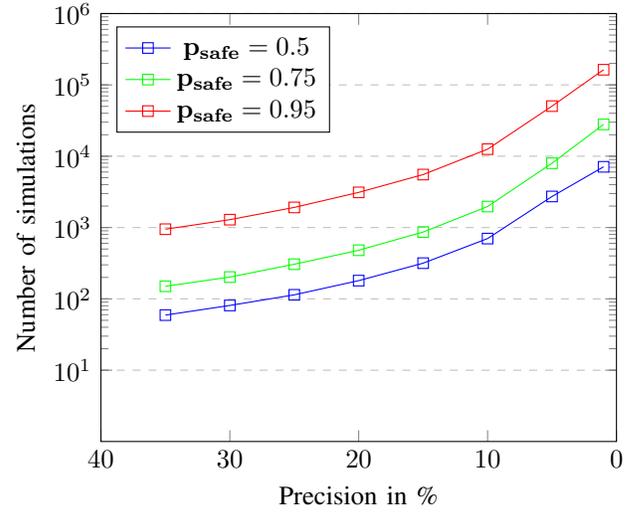
\begin{figure}
\centering
\begin{tikzpicture}[scale=1]
\begin{axis}[
    title={},
    ymode =log,
    xlabel={Precision in \%},
    ylabel={Number of simulations},
    xmin=0, xmax=40,
    ymin=1, ymax=1000000,
    xtick={0, 10, 20, 30, 40},
    ytick={0, 10, 100, 1000, 10000, 100000, 1000000},
    legend pos=north west,
    ymajorgrids=true,
    grid style=dashed,
    x dir=reverse
]

\addplot[color=blue, mark=square]
table [x=precision, y=Probability_0_5, col sep=comma] {number_of_simulations.csv};
\addlegendentry{$\mathbf{p_{safe}}=0.5$} 

\addplot[color=green, mark=square]
table [x=precision, y=Probability_0_75, col sep=comma] {number_of_simulations.csv};
\addlegendentry{$\mathbf{p_{safe}}=0.75$}  

\addplot[color=red, mark=square]
table [x=precision, y=Probability_0_95, col sep=comma] {number_of_simulations.csv};
\addlegendentry{$\mathbf{p_{safe}}=0.95$}  

\end{axis}
\end{tikzpicture}
\caption{Number of simulations required for the brute-force process to reach a desired relative precision, depending on $\mathbf{p_{safe}}$ probability, each point is the average of 100 simulations}
\label{fig:2} 
\end{figure}

In this new process, we sample batches of scenarios. For each input scenario $\bs{x_i}$, the MGP-based proxy provides the probability density function (PDF) $g_i$ of the output $\bs{y_i} \in \mathbb{R}^L$. Integrating this PDF allows us to compute the probability $p_i = \int_{[1, \infty]^L} g_i(y) dy$ for scenario $\bs{x_i}$ to lead to congestion. If the confidence in the outcome is high enough, i.e., if $p_i$ is either very small or very high, we keep the prediction and avoid a simulation. Otherwise, a simulation is performed to obtain the outcome. Fig.3 illustrates the decision mechanism to decide when to keep the prediction or simulate the outcome. At the beginning of the process, only a few outcomes are available to train the proxy. Its quality is poor, and the prediction's confidence is often insufficient to avoid the simulation. Most scenarios are thus simulated. However, the more samples are drawn, the better the proxy becomes. At some point, most predictions will be confident enough to avoid the simulation, resulting in a considerable time gain for the probability estimation. The proxy-based process is detailed in Algorithm 2.

\vspace{0.2cm}

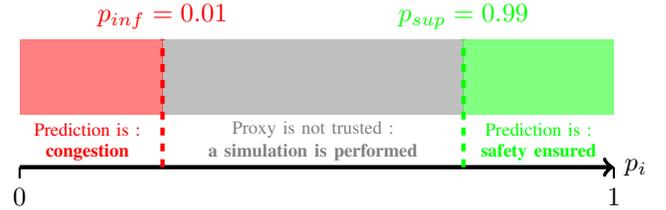
\begin{figure}[H]
\centering
\begin{tikzpicture}
\fill[gray,opacity=0.5] (0,0) -- (4,0) -- (4,1) -- (0,1) -- (0,0);
\fill[red,opacity=0.5] (0,0) -- (0,1) -- (-1.9,1) -- (-1.9,0) -- (0,0);
\fill[green,opacity=0.5] (4,0) -- (4,1) -- (6,1) -- (6,0) -- (4,0);
\draw[->,ultra thick] (-1.9,-0.7)--(6,-0.7) node[right]{$p_i$};
\draw[-,thick] (-1.9,-0.7)--(-1.9,-0.85) node[below]{$0$};
\draw[-,thick] (6,-0.7)--(6,-0.85) node[below]{$1$};
\draw[-, green, ultra thick, dashed] (4,-0.7)--(4,1) node[above]{$p_{sup}=0.99$};
\draw[-, red, ultra thick, dashed] (0,-0.7)--(0,1) node[above]{$p_{inf}=0.01$};
\draw[red, font=\scriptsize] (-1,-0.2) node {Prediction is :};
\draw[red, font=\scriptsize] (-1,-0.5) node {\textbf{congestion}};
\draw[green, font=\scriptsize] (5,-0.2) node {Prediction is : };
\draw[green, font=\scriptsize] (5,-0.5) node {\textbf{safety ensured}};
\draw[gray, font=\scriptsize] (2,-0.2) node {Proxy is not trusted : };
\draw[gray, font=\scriptsize] (2,-0.5) node {\textbf{a simulation is performed}};
\end{tikzpicture}
\caption{Proxy use decision mechanism}
\label{fig:3} 
\end{figure}

We compute the confidence interval using only simulated outcomes in the brute-force process. There is no uncertainty on any $z_i$. In the machine learning process, it is different. When the result is predicted, $z_i$ is no longer known with 100\% confidence because $0 \leq p_i \leq 1$. Such uncertainty must be considered in the confidence interval around the estimation of $\mathbf{p_{safe}}$. The classic version of the CLT must be adapted to our problem. 

\begin{algorithm}[!t]
\caption{Proxy-based process} \label{alg:CV2}
\textbf{Inputs} : 
\begin{itemize}
    \item scenario sampler $\mathbb{X}$
    \item network simulator $\mathcal{S}$
    \item MGP proxy $\Tilde{\mathcal{S}}$
    \item confidence interval precise enough : $CI_{good}?$ tool
    \item confidence interval level $1 - \alpha$
\end{itemize}
\textbf{Output} : $I = [p_{min}, p_{max}]$, $1 - \alpha$ confidence interval level around $\mathbf{p_{safe}}$ \\
\textbf{Hyperparameters} :
\begin{itemize}
    \item Batch size $m_{batch}$
    \item MGP parameters
    \item Proxy usage confidence threshold $p_{inf}$ and $p_{sup}$
\end{itemize}
\begin{algorithmic}[1]
\State Initialize $m=0$ and $I_m = [0, 1]$ 
\State Initialize proxy $\Tilde{\mathcal{S}}$
\While{not $CI_{good}?(I_m)$}
    \State Draw $m_{batch}$ scenarios $\bs{x_1},...,\bs{x_{m_{batch}}}$ using $\mathbb{X}$
    \State Increase $m=m+1$
    \For{$i \in [1, m_{batch}]$}
        \State Compute the probability $p_i = \mathbb{P}(z_i=1) = \Tilde{\mathcal{S}}(\bs{x_i})$
        \If{$p_i \leq P_{min}$ or $p_i \geq P_{max}$}
            \State Draw $w_i \sim B(p_i)$
            \If{$w_i=1$}
                \State Set $q_i = p_i$
            \Else
                \State Set $q_i = 1 - p_i$
            \EndIf
        \Else
            \State Set $w_i = \mathcal{S}(\bs{x_i})$
            \State Set $q_i = 1$
        \EndIf
        \State Add outcome $(w_i, q_i)$ to the bag $\mathcal{B} = \mathcal{B} \cup \{(w_i, q_i)\}$ 
    \EndFor
    \State Update MGP parameters with simulated samples in $\mathcal{B}$
    \State Compute $I_m$ of level $1 - \alpha$ using $\mathcal{B}$
\EndWhile
\end{algorithmic}
\end{algorithm}

\subsection{Central Limit Theorem adaptation}

Let $\bs{x_1},...,\bs{x_m} \in \mathcal{X}$ be $m$ scenarios :

\begin{itemize}
    \item $z_1,...,z_m \in \{0, 1\}$ denote the true answers given by the simulator
    \item $\forall i \in [1, m], \mathbb{P}(z_i = 1) = \mathbf{p_{safe}}$
    \item $w_1,...,w_m \in \{0, 1\}$ denote the answers' prediction 
    \item We define $\forall i \in [1, p], q_i = \mathbb{P}(w_i = z_i)$
\end{itemize}

We suppose that we observe $w_1,...,w_m$ and $q_1,...,q_m$, we look for a confidence interval of level $1 - \alpha$ around $\mathbf{p_{safe}}$. 

\vspace{0.2cm}

\textbf{Proposition 1}

We set :

\begin{align*}
    \overline{Z_m} &= \frac{\sum_{i=1}^m w_i - (1-q_i)}{\sum_{i=1}^m 2q_i - 1} \\
    v_m &= \frac{\left(\sum_{i=1}^m 2q_i-1\right)^2}{\sum_{i=1}^m (2q_i-1)^2} \\
    \sigma_m^2 &= \frac{\sum_{i=1}^m q_i(1-q_i)}{\sum_{i=1}^m (2q_i-1)^2} \\
\end{align*}

Then,

\begin{align*}
    \sqrt{v_m} \frac{\overline{Z_m} - \mathbf{p_{safe}}}{\sqrt{\sigma_m^2 + \mathbf{p_{safe}}(1-\mathbf{p_{safe}})}} \longrightarrow_d \mathcal{N}(0, 1)
\end{align*}

\textit{Proof} : See Annex A

\vspace{0.2cm}

The MGP does not provide us directly $w_i$, nor $\mathbb{P}(w_i = z_i)$, but only $p_i = \mathbb{P}(z_i=1|MGP \; prediction)$. The previous proposition cannot be directly applied. The issue is overcome by generating $w_i$ using a Bernoulli distribution $B(p_i)$ and choosing $q_i = p_i$ if $w_i = 1$ or $q_i = 1 - p_i$ if $w_i = 0$. Then, we have $\mathbb{P}(w_i = z_i) = q_i$ in all cases. All the information the MGP provides is contained in $q_i$, and the random drawings ensure the correctness of the propositions.

\vspace{0.2cm}

If there is no uncertainty on the observations $w_1,...,w_m$, i.e. $\forall i \in [1, p], q_i=1$, then $w_i \sim B(\mathbf{p_{safe}})$ and we retrieve the CLT for the $z_i$. On the contrary, if predictions provide no information, i.e., $\forall i \in [1, p], q_i=0.5$, then $w_i \sim B(1/2)$ and we retrieve the CLT for balanced Bernoulli variables. Our resulting CLT is a combination of these two CLTs. The balance between the two is determined by the $q_i$. Both terms $\overline{Z_p}$ and $v_p$ can be interpreted as the average $\overline{w}$ and $p$ while $\sigma_p^2$ is an additional variance term. It represents a measure of the quantity of uncertainty contained in the observations. The evolution of this term will be studied in the experiments.

\vspace{0.3cm}

\textbf{Proposition 2}

Denoting by $Q_{\alpha}$ the $1-\alpha$ quantile of the standard normal distribution, the asymptotic confidence interval around $\mathbf{p_{safe}}$ is $[p_{min}, p_{max}]$ with :

\begin{align*}
    p_{min} &= \frac{\overline{Z_m} + \frac{Q_{\frac{\alpha}{2}}^2}{2v_m} - Q_{\frac{\alpha}{2}} \sqrt{ \frac{\overline{Z_m}\left(1-\overline{Z_m}\right)}{v_m} +\frac{\sigma_m^2}{v_m} \left[ 1 + \frac{Q_{\frac{\alpha}{2}}^2}{v_m} \right] + \frac{Q_{\frac{\alpha}{2}}^2}{4v_m^2}}}{1 + \frac{Q_{\frac{\alpha}{2}}^2}{v_m}} \\
    p_{max} &= \frac{\overline{Z_m} + \frac{Q_{\frac{\alpha}{2}}^2}{2v_m} + Q_{\frac{\alpha}{2}} \sqrt{ \frac{\overline{Z_m}\left(1-\overline{Z_m}\right)}{v_m} + \frac{\sigma_m^2}{v_m} \left[ 1 + \frac{Q_{\frac{\alpha}{2}}^2}{v_m} \right] + \frac{Q_{\frac{\alpha}{2}}^2}{4v_m^2}}}{1 + \frac{Q_{\frac{\alpha}{2}}^2}{v_m}}
\end{align*}

\textit{Proof} : See Annex B

\section{Computational results}

\subsection{Details on the case study}

\underline{\textbf{Simulator details}}

Both processes' performances are tested on a small network of $L=5$ lines and $N=10$ generator nodes. Flows on the lines are computed using the PTDF \cite{Sosic2014Features}, and $P^{max}$ values, considered as features of the zone. When faraway topological changes occur outside the area, it usually results in modifying the PTDF values. Including topological changes in the scenario $\bs{x}$ would result in an exploding dimension, making the MGP inefficient. Such phenomenon is thus considered a disturbance: at each simulation, we add a small white noise $\epsilon_n$ to the PTDF. Our modeling, however, does not embody significant changes close to the zone, as it would completely redefine the PTDF values. It is one of the limitations of this model. 

The actual NAZA controller cannot be integrated into our simulator. Our scenarios have no temporal dimension, and NAZA actions are based on the estimation of the past flow increase. To model fictitious actions, we thus add a $\beta \in [0, 1]$ coefficient that curtails the available renewable power for all nodes.

\begin{align*}
F_l = \beta \sum_{n=1}^N (PTDF_{l, n} + \epsilon_n) x_n
\end{align*}

$\beta$ is selected based on the minimum required curtailment that avoids congestion for scenarios with similar total power injection in the zone. More  specifically, given a set of scenarios $\bs{x_1},...,\bs{x_J}$, we compute the minimum required curtailment that avoids congestion: $\beta_j = \arg\max\limits_{\beta \in [0,1]} \left\{ \beta \sum_{n=1}^N (PTDF_{l, n} + \epsilon_n) x^{j}_n < 1 \right\}$. Then, for another scenario $\bs{x}$, we randomly draw its curtailment value $\beta$ among the set of candidates : $Cand(\bs{x})= \left\{ \beta_j, j \in [1, J], \left| \sum_{n=1}^N x_n - x^{j}_n \right| < \eta \right\}$, with default value to $1$ if the set is empty. Fig.4 illustrates the $\beta$ selection procedure. Again, the goal of this simple fictitious simulator is only to illustrate the proxy-based process. It will be tested with real network simulators, including a real NAZA automaton.

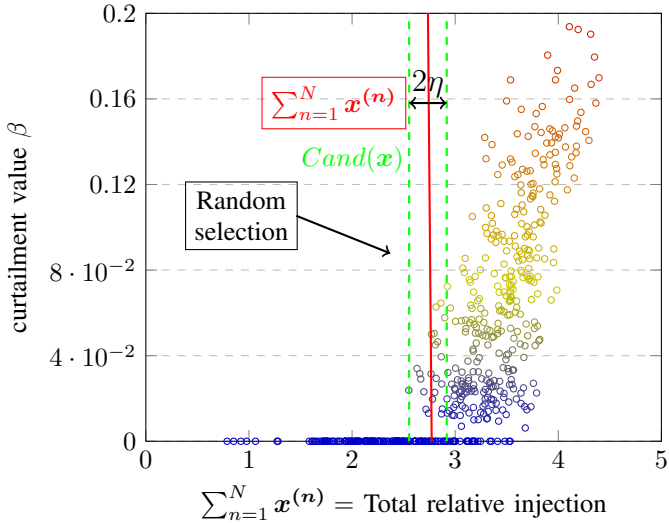
\begin{figure}
\centering
\begin{tikzpicture}[scale=1]
\begin{axis}[
    title={},
    xlabel={$\sum_{n=1}^N \bs{x^{(n)}} = $ Total relative injection},
    ylabel={curtailment value $\beta$},
    xmin=0, xmax=5,
    ymin=0, ymax=0.2,
    xtick={0, 1, 2, 3, 4, 5},
    ytick={0, 0.04, 0.08, 0.12, 0.16, 0.20},
    legend pos=north west,
    ymajorgrids=true,
    grid style=dashed,
]
\addplot+[only marks, scatter, mark=o, mark size=1.2pt]
table[meta=NAZA_curtailment_history]{NAZA_curtailment_history.dat};
\end{axis}

\draw [thick, red] (3.8,0) -- (3.75,5.7);
\node[draw,align=center, red] at (2.5, 4.5) {$\sum_{n=1}^N \bs{x^{(n)}}$};
\draw [thick, dashed, green] (3.5,0) -- (3.5,5.7);
\draw [thick, dashed, green] (4,0) -- (4,5.7);
\draw[<->][thick] (3.5,4.5)--(4,4.5) node{};
\draw[->][thick] (2,3)--(3.25,2.5) node{};
\node[draw,align=center] at (1.27,3) {Random \\ selection};
\draw[black, font=\large] (3.75,4.75) node {$2 \eta$};
\draw[green] (2.75,3.75) node {$Cand(\bs{x})$};

\end{tikzpicture}
\caption{Selection of the curtailment value $\beta$ for a given scenario based on historical curtailment decisions}
\label{fig:4} 
\end{figure}

\vspace{0.2cm}

\underline{\textbf{Convergence criterion}}

We sample random scenarios until a certain number of simulations have been reached. For the brute-force process, the number of simulations matches the number of sampled scenarios. More iterations can be performed for the proxy-based process, depending on the proxy's accuracy.

\vspace{0.35cm}

\underline{\textbf{Computational details}}
Over the iterations, we might have tens of thousands of simulated scenarios to consider when training the MGP. It, however, implies inverting a covariance matrix with a dimension equal to the number of simulated scenarios, thus over tens of thousands. It is too time- and space-consuming to be efficiently computed. Nevertheless, when computing the likelihood of a new scenario, many previously simulated scenarios will be very different. Their correlation with the new input will be negligible. All these scenarios could thus be neglected when estimating the parameters of the conditional distribution. Inspired by \cite{Liu2020When}, we consider only a small proportion of the simulated scenarios when computing the likelihood of a new input $\bs{x_0}$.

\begin{itemize}
	\item At most $N^*$ scenarios
	\item Only neighbors scenarios that are at a distance less than $d^*$: $d(\bs{x}, \bs{x_0}) < d^*$
\end{itemize}

The last point that deserves to be highlighted is the computation of the integral for $p_i$. Both processes are implemented in Python; we use the Scipy library \cite{Virtanen2020Scipy} to approximate the integral numerically. In small dimensions, the evaluation of the integral is fast. In higher dimensions, the time quickly increases, cf Fig.5. It is another limitation to overcome to generalize this approach to bigger networks.

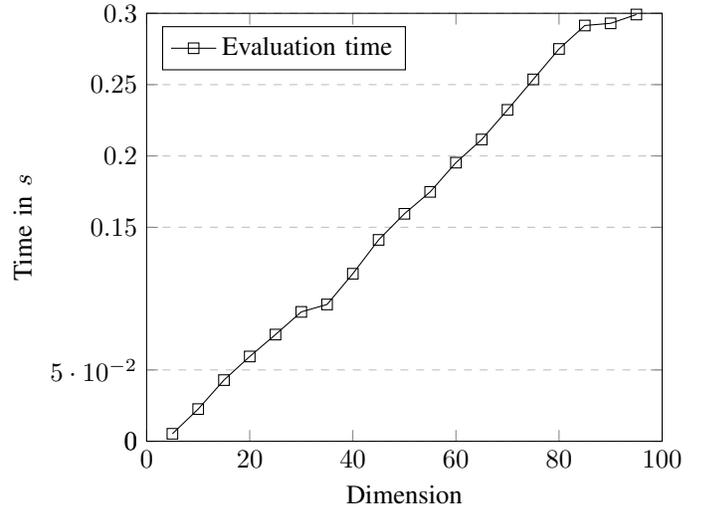
\begin{figure}
\centering
\begin{tikzpicture}[scale=1]
\begin{axis}[
    title={},
    xlabel={Dimension},
    ylabel={Time in $s$},
    xmin=0, xmax=100,
    ymin=0, ymax=0.3,
    xtick={0, 20, 40, 60, 80, 100},
    ytick={0, 0.05, 0,1, 0.15, 0.2, 0.25, 0.3},
    legend pos=north west,
    ymajorgrids=true,
    grid style=dashed,
]

\addplot[color=black, mark=square]
table [x=dimension, y=CDF_computation_time, col sep=comma] {CDF_computation_time.csv};
\legend{Evaluation time}   

\end{axis}
\end{tikzpicture}
\caption{Integral evaluation time depending on the dimension, 10000 averaged simulations per point}
\label{fig:5} 
\end{figure}

\subsection{Proxy performances}

In this subsection, we illustrate how the MGP manages to learn the behavior of the black-box function to act as a proxy of the simulator. We plot three different indicators.

First, we display in Fig.6 the evolution of $\sigma_*$ across the process's iterations. $\sigma_*$ is the scale of the covariance matrix of the posterior distribution. A small  $\sigma_*$ reflects a low variance and thus a high confidence in the predicted outcome. $\sigma_*$ entirely depends on the point we wish to predict, rendering the plot irregular. To catch the global trend, we display a moving average.

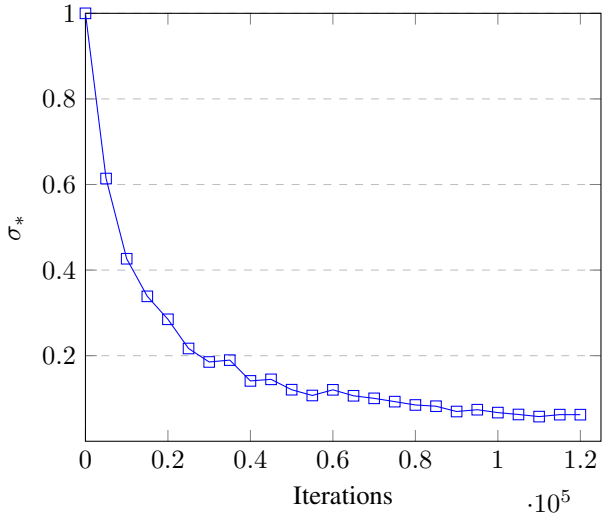
\begin{figure}
\centering
\begin{tikzpicture}[scale=1]
\begin{axis}[
    title={},
    xlabel={Iterations},
    ylabel={$\sigma_*$},
    xmin=0, xmax=125000,
    ymin=0, ymax=1,
    xtick={0, 20000, 40000, 60000, 80000, 100000, 120000},
    ytick={0.2, 0.4, 0.6, 0.8, 1},
    legend pos=north west,
    ymajorgrids=true,
    grid style=dashed,
]

\addplot[color=blue, mark=square]
table [x=iteration, y=sigma_star, col sep=comma] {sigma_star.csv};  

\end{axis}
\end{tikzpicture}
\caption{Evolution of the 200 mobile average of $\sigma_*$ during the proxy-based process}
\label{fig:6} 
\end{figure}

In the beginning, the MGP has no observations from which to learn. The variance of the posterior distribution is $\sigma_*=\sigma_f + \sigma_0=1$, an arbitrary prior choice. Throughout the process, we simulate more scenarios, increasing the chances of having many correlated points close to the new sampled observation. Mathematically, this boils down to a higher value for the Mahalanobis term $\bs{\Sigma_{[1, m], m+1}}^T \bs{\Sigma_m}^{-1} \bs{\Sigma_{[1,m], m+1}}^T$, and thus a lower variance. The decrease of $\sigma_*$ illustrates the increasing density of simulated scenarios in the information structure, directly linked to a downward variance posterior distribution and a high confidence in the prediction. $\sigma_*$ could amount to a loss function, whose decrease attests that our MGP indeed learns the simulator's answer.

The second indicator is related to the evolution of confidence in the outcome's prediction across the process's iterations. We plot in Fig.7 the evolution of the prediction's entropy: $H(z_i) = \frac{-p_i log(p_i) - (1-p_i) log(1-p_i)}{log(2)}$. A small entropy characterizes a probability close to $0$ or $1$, thus a confident prediction, while a higher entropy stands for a more uncertain prediction. Again, we display a moving average to avoid a very irregular plot.

\begin{figure}
\centering
\begin{tikzpicture}[scale=1]
\begin{axis}[
    title={},
    xlabel={Iterations},
    ylabel={Entropy},
    xmin=0, xmax=125000,
    ymin=0, ymax=1,
    xtick={0, 20000, 40000, 60000, 80000, 100000, 120000},
    ytick={0.2, 0.4, 0.6, 0.8, 1},
    legend pos=north west,
    ymajorgrids=true,
    grid style=dashed,
]

\addplot[color=blue, mark=square]
table [x=iteration, y=entropy, col sep=comma] {entropy.csv};

\end{axis}
\end{tikzpicture}
\caption{Evolution of the 200 mobile average of the proxy's prediction entropy during the proxy-based process
}
\label{fig:7} 
\end{figure}
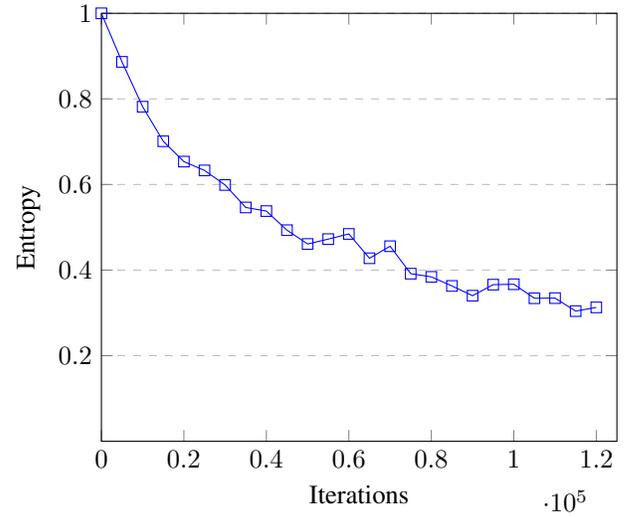

In the beginning, the posterior distribution is $\mathcal{N}(0, \bs{\Omega})$. $\bs{\Omega}$ scale is adapted so that the prior probability is unbiased: $p_1=q_1=\frac{1}{2}$ and $H(z_1)=1$. The congestion probability is mainly influenced by whether or not $\mu_*$ is close, relatively to $\sigma_*$, to the congestion limit, i.e., the boundaries of the $[0,1]^N$ volume. As $\sigma_*$ decreases, it becomes less and less likely to be close to the border. Thus, the computed probabilities are increasingly extreme, close to $0$ (confidence in an overload) or very close to $1$ (confidence in a safe line).

Finally, the last interesting factor to display is the \% of simulated scenarios since the beginning of the certification process. The lower the \% is, the more simulations were avoided, and thus the better the proxy is.

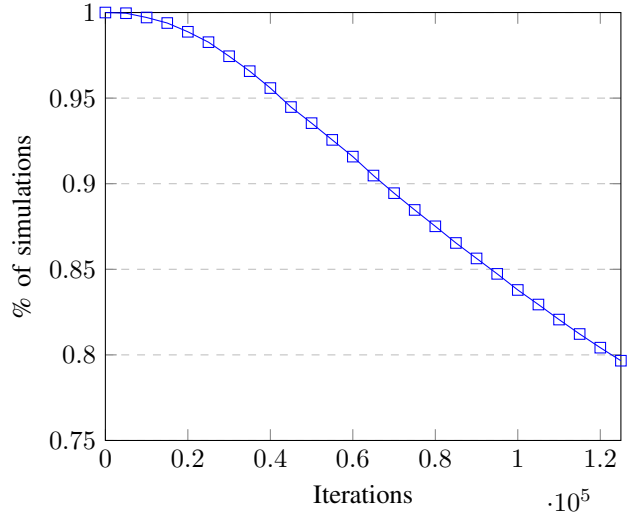
\begin{figure}
\centering
\begin{tikzpicture}[scale=1]
\begin{axis}[
    title={},
    xlabel={Iterations},
    ylabel={\% of simulations},
    xmin=0, xmax=125000,
    ymin=0.75, ymax=1,
    xtick={0, 20000, 40000, 60000, 80000, 100000, 120000},
    ytick={0.75, 0.8, 0.85, 0.9, 0.95, 1},
    legend pos=north west,
    ymajorgrids=true,
    grid style=dashed,
]

\addplot[color=blue, mark=square]
table [x=iteration, y=perc_of_simulations, col sep=comma] {perc_of_simulations.csv};

\end{axis}
\end{tikzpicture}
\caption{Evolution of the total \% of simulated scenarios since the beginning during the proxy-based process}
\label{fig:8} 
\end{figure}

Initially, the proxy's confidence is insufficient to replace an accurate simulation, and most scenarios are simulated. Then, as the proxy quality improves and the prediction's entropy decreases, the proxy is used more and more often. Over the last $1000$ of sampled scenarios, less than $50\%$ are simulated, leading to $20\%$ of avoided simulations.

\subsection{Processes performances}

To conclude the experiments section, we present the performances of each process for probability estimation. Given a maximum amount of simulations, we compare in Table 1 the estimated $\mathbf{p_{safe}}$ probability and the confidence interval obtained by both processes. 

The brute-force process performs 100,000 iterations and reaches a relative error of 0.05\%. Comparatively, the proxy-based process gets over 125,000 iterations and avoids more than 25,000 simulations. The proxy-based confidence interval is computed with more iterations, leading to a smaller relative error and a smaller length. The performance gap is almost $20\%$ above the brute-force process. Since the proxy is continuously enhanced across the simulations, the gap between the two processes would be even more significant for a higher maximum amount of simulations.

\begin{table}
\centering
\begin{tabular}{|p{2.3cm}||p{2.3cm}|p{2.3cm}|}
 \hline
 \multicolumn{3}{|c|}{\cellcolor{blue!20} $\mathbf{p_{safe}} = 0.9044$} \\
 \hline
 \cellcolor{black} & \cellcolor{red!25} Brute-force & \cellcolor{green!25} Proxy-based\\
 \hline
 \hline
 Number of iterations & \cellcolor{red!15} 125575 & \cellcolor{green!15} 100000 \\
 \hline
 $p_{min}$ & \cellcolor{red!15} 0.9018 & \cellcolor{green!15} 0.9023 \\
 \hline
 $p_{max}$ & \cellcolor{red!15} 0.9067 & \cellcolor{green!15} 0.9064 \\
 \hline
 Relative error & \cellcolor{red!15} 0.05\% & \cellcolor{green!15} 0.041\% \\
 \hline
 Confidence interval length & \cellcolor{red!15} 0.0049 & \cellcolor{green!15} 0.0041 \\
 \hline
\end{tabular}
\caption{Processes performances comparison}
\label{table:1}
\end{table}

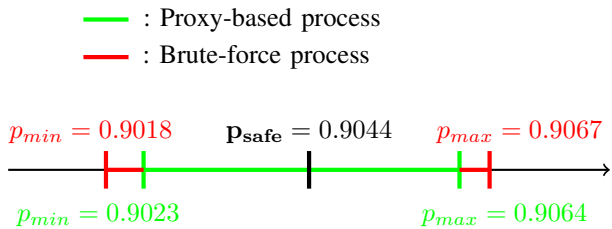
\begin{figure}[H]
\centering
\begin{tikzpicture}
\draw[->, thick] (0,0)--(8,0);
\draw[-, green, ultra thick] (1.8,-0.25)--(1.8,0.25);
\draw[-, green, ultra thick] (6,-0.25)--(6,0.25);
\draw[-, green, ultra thick] (1.8,0)--(6,0);
\draw[black] (4,0.5) node {$\mathbf{p_{safe}}=0.9044$};
\draw[green] (1.2,-0.6) node {$p_{min}=0.9023$};
\draw[green] (6.6,-0.6) node {$p_{max}=0.9064$};
\draw[-, red, ultra thick] (1.3,-0.25)--(1.3,0.25);
\draw[-, red, ultra thick] (6.4,-0.25)--(6.4,0.25);
\draw[-, red, ultra thick] (1.3,0)--(1.8,0);
\draw[-, red, ultra thick] (6,0)--(6.4,0);
\draw[-, ultra thick] (4,-0.25)--(4,0.25);
\draw[red] (1.1,0.5) node {$p_{min}=0.9018$};
\draw[red] (6.8,0.5) node {$p_{max}=0.9067$};
\draw[-, red, ultra thick] (1,1.5)--(1.6,1.5);
\draw[-, green, ultra thick] (1,2)--(1.6,2);
\draw[black] (3.3,1.5) node {: Brute-force process};
\draw[black] (3.375,2) node {: Proxy-based process};
\end{tikzpicture}
\caption{Confidence intervals obtained by both processes}
\label{fig:9} 
\end{figure}

\section{Conclusion}

In this work, we proposed two processes to estimate the probability of observing congestion on a network. The network and its simulator are a black-box function that outputs the maximum flow observed on each line for a renewable energy production scenario. The brute-force process draws and simulates scenarios until it reaches a satisfying probability estimation. As power system simulations are often very long, they are time-consuming. The proxy-based process trains a Multivariate Gaussian Process to predict the black box's answer, with an exact confidence interval on the prediction. The simulation is avoided when the proxy is confident enough to make the prediction. For a given budget of simulations, it leads to a much more accurate probability estimation, with an error reduced by $20\%$. Many simulations are not performed, allowing the process to run more iterations. The MGP manages here to efficiently learn the multivariate answer of a black-box function with correlated components while providing explicability and theoretical guarantees about the prediction's error. Multivariate Gaussian processes are of great interest for approximating power systems behaviors and can be applied to many problems.

We supposed in this work that the output's correlation matrix $\Omega$ was known, while in practice, it has to be learned. A first extension of this work will be to propose an estimation method of this matrix during the process. Also, the kernel choice can be discussed. This paper aimed to illustrate the proxy-based certification process on a simple black box involving low-dimensional inputs. We aim to extend it to more realistic simulations with a real network and complex scenarios.

\section*{Annex A: Proof of proposition 1}

The proof relies on the Lyapunov version of the CLT \cite{Nowak2015Generalized}. It states that for $m$ given random independent variables  $X_1,..,X_m$, with expected value $\mu_i$ and variance $\sigma_i^2$, if for some $\delta > 0$, Lyapunov condition is satisfied : 

\begin{align*}
    s_m^2 &= \sum_{i=1}^m \sigma_i^2 \\
    \lim\limits_{\substack{m \to \infty}}& \frac{1}{s_m^2} \sum_{i=1}^m \mathbb{E} \left[ \left| X_i - \mu_i \right|^{2 + \delta} \right] = 0
\end{align*}

Then, 

\begin{align*}
    \frac{1}{s_m} \sum_{i=1}^m (X_i - \mu_i) \longrightarrow_d \mathcal{N}(0, 1)
\end{align*}

In our case, one can easily check that $\forall i \in [1, m]$ :

\begin{align*}
    w_i &\sim B\left((2 q_i - 1) \mathbf{p_{safe}} + (1 - q_i)\right) \\
    \mathbb{E} \left[ w_i \right] &= (2 q_i - 1) \mathbf{p_{safe}} + (1 - q_i) \\
    V(w_i) &= q_i(1-q_i) + \mathbf{p_{safe}} (1 - \mathbf{p_{safe}}) (2 q_i - 1)^2
\end{align*}

Let $A_1,...,A_m$ be $m$ random independant variables following the Bernoulli distributions $B(a_1),...,B(a_m)$, then :

\begin{align*}
    s_m^2 &= \sum_{i=1}^m a_i(1-a_i) \\
    \mathbb{E} \left[|A_i-\mu_i|^{2+\delta}\right] &= a_i(1-a_i)^{2+\delta} + (1-a_i)a_i^{2+\delta} \\
    &\leq 2a_i(1-a_i)
\end{align*}

Since we suppose that the simulator's answer is non-deterministic, there is always a residual uncertainty on the result's prediction. It implies that $q_i(1-q_i)$ does not converge toward $0$ and thus the serie $\sum_{i=1}^m q_i(1-q_i)$ diverges toward $+ \infty$. We deduce that :

\begin{align*}
\lim\limits_{\substack{m \to \infty}} \sum_{i=1}^m \frac{\mathbb{E} \left[|A_i-\mu_i|^{2+\delta}\right]}{s_p^{2+\delta}} &\leq \lim\limits_{\substack{m \to \infty}} \frac{2 \sum_{i=1}^m a_i(1-a_i)}{\left(\sum_{i=1}^m a_i(1-a_i)\right)^{1+\frac{\delta}{2}}} \\
&= 0
\end{align*}

Lyapunov's condition is thus verified for our set of Bernoulli variables. We apply the theorem to $W_1,..., W_m$: 

\begin{align*}
      &\frac{\sum_{i=1}^m W_i - (2q_i-1)\mathbf{p_{safe}} - (1-q_i)}{\sqrt{\sum_{i=1}^m q_i(1-q_i) + \mathbf{p_{safe}}(1-\mathbf{p_{safe}})(2q_i-1)^2}} \\
    = &\frac{\sum_{i=1}^m W_i - (1-q_i) - \mathbf{p_{safe}} \sum_{i=1}^m (2q_i-1)}{\sqrt{\sum_{i=1}^m q_i(1-q_i) + \mathbf{p_{safe}}(1-\mathbf{p_{safe}}) \sum_{i=1}^m (2q_i-1)^2}} \\
    = & \sqrt{\frac{\left(\sum_{i=1}^m (2q_i-1)\right)^2}{\sum_{i=1}^m (2q_i-1)^2}} \frac{\frac{\sum_{i=1}^m W_i - (1-q_i)}{\sum_{i=1}^m (2q_i-1)} - \mathbf{p_{safe}}}{\sqrt{\frac{\sum_{i=1}^m q_i(1-q_i)}{\sum_{i=1}^m (2q_i-1)^2} + \mathbf{p_{safe}}(1-\mathbf{p_{safe}})}} \\
    = & \sqrt{v_m} \frac{\overline{Z_m}-\mathbf{p_{safe}}}{\sqrt{\sigma_m^2 + \mathbf{p_{safe}}(1-\mathbf{p_{safe}})}}
\end{align*}

\section*{Annex B: Proof of proposition 2}

Asymptotically, we know that $\sqrt{v_m} \frac{\overline{Z_m} - \mathbf{p_{safe}}}{\sqrt{\sigma_m^2 + \mathbf{p_{safe}}(1-\mathbf{p_{safe}})}} \sim \mathcal{N}(0, 1)$. To obtain the result, we solve the following inequality :

\begin{align*}
    &-q_{\frac{\alpha}{2}} \leq \sqrt{v_m} \frac{\overline{Z_m} - \mathbf{p_{safe}}}{\sqrt{\sigma_m^2 + \mathbf{p_{safe}}(1-\mathbf{p_{safe}})}} \leq q_{\frac{\alpha}{2}} \\
    & \implies v_m \frac{\left(\overline{Z_m} - \mathbf{p_{safe}}\right)^2}{\sigma_m^2 + \mathbf{p_{safe}}(1-\mathbf{p_{safe}})} \leq q_{\frac{\alpha}{2}}^2 \\
    & \implies \overline{Z_m}^2 + \mathbf{p_{safe}}^2 - 2\mathbf{p_{safe}} \overline{Z_m} \leq \sigma_m^2 \frac{q_{\frac{\alpha}{2}}^2}{v_m} + \mathbf{p_{safe}} \frac{q_{\frac{\alpha}{2}}^2}{v_m} \\ 
    &- \mathbf{p_{safe}}^2 \frac{q_{\frac{\alpha}{2}}^2}{v_m} \\
    & \implies \left[ 1 + \frac{q_{\frac{\alpha}{2}}^2}{v_m} \right]\mathbf{p_{safe}}^2 + \left[ -2 \overline{Z_m} - \frac{q_{\frac{\alpha}{2}}^2}{v_m} \right]\mathbf{p_{safe}} \\
    &+ \left[ \overline{Z_m}^2 - \sigma_m^2 \frac{q_{\frac{\alpha}{2}}^2}{v_m} \right] \leq 0 \\
    \Delta &= \left[ 2 \overline{Z_m} + \frac{q_{\frac{\alpha}{2}}^2}{v_m} \right]^2 - 4 \left[ 1 + \frac{q_{\frac{\alpha}{2}}^2}{v_m} \right] \left[ \overline{Z_m}^2 - \sigma_m^2 \frac{q_{\frac{\alpha}{2}}^2}{v_m} \right] \\
    &= 4 \overline{Z_m}^2 + \frac{q_{\frac{\alpha}{2}}^4}{v_m^2} + 4 \overline{Z_m} \frac{q_{\frac{\alpha}{2}}^2}{v_m} - 4 \overline{Z_m}^2 + 4 \sigma_m^2 \frac{q_{\frac{\alpha}{2}}^2}{v_m} \\
    &- 4 \overline{Z_m}^2 \frac{q_{\frac{\alpha}{2}}^2}{v_m} + 4 \sigma_m^2  \frac{q_{\frac{\alpha}{2}}^4}{v_m^2} \\ 
    &= 4 \frac{q_{\frac{\alpha}{2}}^2}{v_m} \left[ \overline{Z_m}\left(1-\overline{Z_m}\right) + \sigma_m^2 \left[ 1 + \frac{q_{\frac{\alpha}{2}}^2}{v_m} \right] + \frac{q_{\frac{\alpha}{2}}^2}{4v_m} \right]\\
\end{align*}

We can finally derive the expressions of $p_{min}$ and $p_{max}$ :

\begin{align*}
    p_{min} &= \frac{\overline{Z_m} + \frac{q_{\frac{\alpha}{2}}^2}{2v_m} - q_{\frac{\alpha}{2}} \sqrt{ \frac{\overline{Z_m}\left(1-\overline{Z_m}\right)}{v_m} + \frac{\sigma_m^2}{v_m} \left[ 1 + \frac{q_{\frac{\alpha}{2}}^2}{v_m} \right] + \frac{q_{\frac{\alpha}{2}}^2}{4v_m^2}}}{1 + \frac{q_{\frac{\alpha}{2}}^2}{v_m}} \\
    p_{max} &= \frac{\overline{Z_m} + \frac{q_{\frac{\alpha}{2}}^2}{2v_m} + q_{\frac{\alpha}{2}} \sqrt{ \frac{\overline{Z_m}\left(1-\overline{Z_m}\right)}{v_m} + \frac{\sigma_m^2}{v_m} \left[ 1 + \frac{q_{\frac{\alpha}{2}}^2}{v_m} \right] + \frac{q_{\frac{\alpha}{2}}^2}{4v_m^2}}}{1 + \frac{q_{\frac{\alpha}{2}}^2}{v_m}} \\
\end{align*}

\end{document}